\documentclass[a4paper,fleqn,usenatbib]{mnras}
\usepackage{amsmath}
\usepackage{epsfig} 
\usepackage{graphicx}	
\usepackage{amssymb}	
\newcommand{\be}{\begin{equation}}
\newcommand{\e}{\end{equation}}
\newcommand{\bear}{\begin{eqnarray}}
\newcommand{\ear}{\end{eqnarray}}

\newcommand{\hmpc}{{\, h^{-1}\, {\rm Mpc}}}

\def\aj{AJ}
\def\apj{ApJ}
\def\apjs{ApJS}
\def\jcap{JCAP}
\def\mnras{MNRAS}
\def\aap{A\&A}

\def\nat{Nature}      
\def\apjs{ApJS}
\def\apjl{ApJ Letters}

\title[Bias from anisotropy] {Can anisotropy in the galaxy
  distribution tell the bias?}
    
\author[Pandey, B.]  { Biswajit Pandey\thanks{E-mail:
    biswap@visva-bharati.ac.in} \\ Department of Physics,
  Visva-Bharati University, Santiniketan, Birbhum, 731235, India\\ }
   
 \date{\today}

 \pubyear{2016}
  
\begin{document}
\label{firstpage}
\pagerange{\pageref{firstpage}--\pageref{lastpage}}      
\maketitle
       
\begin{abstract}

We use information entropy to analyze the anisotropy in the mock
galaxy catalogues from dark matter distribution and simulated biased
galaxy distributions from $\Lambda$CDM N-body simulation. We show that
one can recover the linear bias parameter of the simulated galaxy
distributions by comparing the radial, polar and azimuthal
anisotropies in the simulated galaxy distributions with that from the
dark matter distribution. This method for determination of the linear
bias requires only $O(N)$ operations as compared to $O(N^{2})$ or at
least $O(N \log N)$ operations required for the methods based on the
two-point correlation function and the power spectrum. We apply this
method to determine the linear bias parameter for the galaxies in the
2MASS Redshift Survey (2MRS) and find that the 2MRS galaxies in the
$K_{s}$ band have a linear bias of $\sim 1.3$.

\end{abstract}


\section{Introduction}
The homogeneity and isotropy of the Universe on large scales is a
fundamental tenet of modern cosmology. Our current understanding of
the cosmos relies heavily on this principle. Presently a large number
of cosmological observations such as the temperature of the Cosmic
Microwave Background (CMBR) \citep{smoot, fixsen}, X-ray background
\citep{wu,scharf}, angular distributions of radio sources
\citep{wilson,blake}, Gamma-ray bursts \citep{meegan,briggs},
supernovae \citep{gupta,lin}, galaxies \citep{marinoni, alonso} and
neutral hydrogen \citep{hazra} are known to favour the assumption of
statistical isotropy. But this assumption does not hold on small
scales and the anisotropies present on these scales can tell us a lot
about the Universe. For example, the CMBR is not completely isotropic
and the anisotropies imprinted in the CMBR perhaps provide the richest
source of information in cosmology \citep{adeplanck3}. In the current
paradigm, the large scale structures in the Universe are believed to
emerge from the gravitational amplification of the miniscule density
fluctuations generated in the early Universe. The anisotropies in the
CMBR shed light on the conditions that prevailed in the early Universe
whereas the anisotropies in the present day mass distribution help us
to unravel the formation and evolution of the large scale structures
in the Universe.

Currently there exist a wide variety of statistical tools to quantify
the distribution of matter in the Universe. Besides there use in the
study of CMBR anisotropies, the two-point correlation function and the
power spectrum also remain the most popular choice for the study of
clustering. The two-point correlation function \citep{peeb80} measures
the amplitude of galaxy clustering as a function of scale whereas the
shape and amplitude of the power spectrum also provide the information
about the amount and nature of matter in the Universe. The three-point
correlation function and the bispectrum has been also widely used in
the study of clustering in the galaxy distribution. These statistics
are popular as one can directly relate them to the theories of
structure formation.

The distribution of galaxies are believed to trace the mass
distribution on large scales, where the density fluctuations in
galaxies and mass are assumed to be related linearly
\citep{kaiser84,dekel}. In the linear bias assumption, both the two
point correlation function and power spectrum can be employed to
determine the linear bias between galaxies and mass
\citep{nor,teg,zehavi10}. The distribution of the galaxies are
inferred from their redshifts. The peculiar velocities induced by the
density fluctuations perturb their redshifts. This distorts the
clustering pattern of galaxies in redshift space and cause the
two-point correlation function and power spectrum to be
anisotropic. They are suppressed on small scales due to the motion of
galaxies inside virialized structures and enhanced on large scales due
to coherent flows into over dense regions and out of under dense
regions. The anisotropies in the two-point correlation function and
the power spectrum can be decomposed into different angular moments
\citep{kaiser,hamilton} and their ratios can be used to determine the
linear distortion parameter $\beta \approx \frac{\Omega_{m}^{0.6}}{b}$
where $\Omega_{m}$ is the mass density parameter and $b$ is the linear
bias parameter. This method has been used to determine the linear bias
\citep{haw,teg}. One can also use the three-point correlation function
and bispectrum \citep{feldman,verde,gaztanaga} to measure the bias. It
may be noted that some sort of parameter degeneracies are involved in
all these methods. Computing the correlation functions and the poly
spectra are also computationally expensive for very large data sets.

The information entropy is related to the higher order moments of a
distribution and hence captures more information about the
distribution. \citet{pandey16b} propose a method based on the
information entropy-mass variance relation to determine the large
scale linear bias from galaxy redshift surveys. We investigate if this
relation also holds for the anisotropy measure proposed in
\citet{pandey16a} and can one exploit this relation to measure the
linear bias by directly measuring the anisotropy in the galaxy
distribution. An important advantage of this method is the fact that
for any given data set, it is computationally less expensive than the
methods which are based on the two-point correlation function and the
power spectrum. The only disadvantage of the method is that the
information entropy is sensitive to binning and sampling. But this
relative character of entropy does not pose any problem provided the
distributions are compared with the same binning and sampling rate.

The modern redshift surveys (SDSS, \citealt{york}; 2dFGRS,
\citealt{colles}; 2MRS, \citealt{huchra}) have now mapped the galaxy
distribution in the local Universe with unprecedented accuracy. The
SDSS and 2dFGRS are deeper than 2MRS but they only cover parts of the
sky. Moreover the 2MASS redshift survey (2MRS) maps the galaxies over
nearly the entire sky ($\sim 91\%$) out to a distance of $300 \, {\rm
  Mpc}$. Unlike its optical counterparts, the 2MRS selects galaxies in
the near infrared wavelengths around $2 \mu m$ which makes it less
susceptible to extinction and stellar confusion. The old stellar
populations which are otherwise missed by the optical surveys are also
retained in 2MRS due to its operation in the infrared window. The
survey is $97\%$ complete down to the limiting magnitude of
$K_s=11.75$ which provides a fair representation of the mass
distribution in the local Universe. These advantages offered by the
2MRS over the other surveys make it most suitable for the analysis in
the present work.

We use a $\Lambda$CDM model with $\Omega_{m0}=0.31$,
$\Omega_{\Lambda0}=0.69$ and $h=1$ for converting redshifts to
distances throughout our analysis.

A brief outline of the paper follows. In section 2 we describe the
method of analysis followed by a description of the data in section
3. We present the results and conclusions in section 4.


\section{METHOD OF ANALYSIS}

The information entropy was first introduced by Claude Shannon
\citep{shannon48} to find the most efficient way to transmit
information through a noisy communication channel. It quantifies the
uncertainty in the measurement of a random variable. Given a
probabilistic process with probability distribution $p(x)$ where the
random variable $x$ has $n$ outcomes given by $\{x_{i}:i=1,....n\}$,
the average amount of information to describe the random variable $x$
is given by,
\begin{equation}
H(x) =  - \sum^{n}_{i=1} \, p(x_{i}) \, \log \, p(x_{i})
\label{eq:shannon1}
\end{equation}
The quantity $H(x)$ is known as the information entropy of the random
variable $x$.

\citet{pandey16a} propose an anisotropy measure based on the
information entropy and carry out tests on various isotropic and
anisotropic distributions to find that it can efficiently recover
various types of anisotropies inputted in a distribution. The method
divides the entire sky into equal area pixels by carrying out uniform
binning of $\cos\theta$ and $\phi$. Here $\theta$ and $\phi$ are
respectively the polar and azimuthal angles in spherical polar
co-ordinates. The entire sky is divided into
$m_{total}=m_{\theta}m_{\phi}$ angular bins or pixels where
$m_{\theta}$ and $m_{\phi}$ correspond to the number of bins used for
binning $\cos\theta$ and $\phi$ respectively. At any distance $r$,
each of these pixels subtend equal volumes. The method counts the
number of galaxies inside each of these volume elements and define a
random variable $X_{\theta\phi}$ with $m_{total}$ outcomes each given
by, $f_{i}=\frac{n_{i}(<r)}{\sum^{m_{total}}_{i=1} \, n_{i}(<r)}$. The
$f_{i}$ represents the probability of finding a randomly selected
galaxy in the $i^{th}$ bin and we have $\sum^{m_{total}}_{i=1} \,
f_{i}=1$.

One can write the information entropy associated with $X_{\theta\phi}$
for a given $r$ as,
\begin{eqnarray}
H_{\theta\phi}(r)& = &- \sum^{m_{total}}_{i=1} \, f_{i}\, \log\, f_{i} \nonumber\\ &=& 
\log N - \frac {\sum^{m_{total}}_{i=1} \, n_i (<r)\, \log n_i(<r)}{N}
\label{eq:shannon2}
\end{eqnarray}
where $N$ is the total number of galaxies which are distributed within
a distance $r$ across all the bins. The base of the logarithm can be
chosen arbitrarily and we choose it to be $10$ for the present work.

The information entropy $H_{\theta\phi}$ will have the maximum value
$(H_{\theta\phi})_{max}=\log \, m_{total}$ for a given choice of
$m_{\theta}$, $m_{\phi}$ and $r$ when the probability $f_{i}$ becomes
$\frac{1}{m_{total}}$ and identical for all the bins.  The anisotropy
parameter $a_{\theta\phi}(r)$ is defined as
$a_{\theta\phi}(r)=1-\frac{H_{\theta\phi}(r)}{(H_{\theta\phi})_{max}}$.
Ideally an isotropic distribution will always have maximum entropy and
consequently $a_{\theta\phi}(r)$ will be zero. The value of
$a_{\theta\phi}(r)$ thus characterizes the degree of anisotropy
present in a distribution. It may be noted that a discrete isotropic
distribution will always show a small but non-zero value for
$a_{\theta\phi}(r)$ due to shot noise. In other words, the measure is
sensitive to binning and sampling and one should always compare the
degree of anisotropy in two different distributions with same binning
and sampling rate.

Besides characterizing the radial anisotropy $a_{\theta\phi}(r)$, one
can also quantify the polar and azimuthal anisotropies by measuring
$a_{\phi}(\theta)=1-\frac{H_{\phi}}{(H_{\phi})_{max}}$ and
$a_{\theta}(\phi)=1-\frac{H_{\theta}}{(H_{\theta})_{max}}$
respectively.  This would requite to carry out the sum in
\autoref{eq:shannon2} respectively over $m_{\phi}$ or $m_{\theta}$
bins instead of $m_{total}$. We fix the radius to a value $r_{max}$ in
this case. The number $N$ in these cases would be the total number of
galaxies residing in the $m_{\phi}$ or $m_{\theta}$ bins available at
different $\theta$ or $\phi$ respectively.

One can write the number counts as $n_{i}(<r)=n_{0}(<r)+\delta
n_{i}(<r)$ where $\delta n_{i}(<r)$ are small fluctuations around the
mean $n_{0}(<r)$ and express the entropy deficit
$(H_{\theta\phi})_{max}-H_{\theta\phi}(r)$ in terms of the different
moments of the distribution \citep{pandey16b} as,
\begin{eqnarray}
 (H_{\theta\phi})_{max}-H_{\theta\phi}(r) \,=\,
  \frac{1}{2\,m_{total}\,n_{0}(<r)^{2}} \sum^{m_{total}}_{i=1}\delta
  n_i^{2}(<r) \nonumber \\ - \frac{1}{6\,m_{total}\,n_{0}(<r)^{3}}
  \sum^{m_{total}}_{i=1}\delta n_i^{3}(<r) \nonumber \\ +
  \frac{1}{3\,m_{total}\,n_{0}(<r)^{4}} \sum^{m_{total}}_{i=1}\delta
  n_i^{4}(<r)-...
\label{eq:shannon3}
\end{eqnarray}

The first term in the above expression can be clearly identified with
the variance $\sigma^{2}_{r}$ in the number counts. Neglecting the
contributions from the higher order moments in the limit $\frac{\delta
  n_{i}(<r)}{n_{0}(<r)}<<1$, one can relate the entropy deficit to the
variance in number counts as,
\begin{eqnarray} 
(H_{\theta\phi})_{max}-H_{\theta\phi}(r) \,=\,
  \frac{\sigma^{2}_{r}}{2}
\label{eq:shannvar}
\end{eqnarray}

One may note that if the particles or galaxies are assumed to have
equal masses then this variance in the number counts can be treated as
the mass variance in those volume elements. The cosmological mass
variance of a smoothed density field can be also determined from the
power spectrum as,
\begin{eqnarray} 
\sigma^{2}_{r}\,=\,
  \frac{1}{(2\,\pi)^{2}}\, \int_{0}^{\infty} k^{2} P(k)\widetilde{W}^{2}(kr) dk
    \label{eq:variance2}
\end{eqnarray}
where, $r$ is the size of the filter used for smoothing, $P(k)$ is the
power spectrum and $\widetilde W(kr)$ is the Fourier transform of the
filter. The filter shape has to be specified which would then
determine $\widetilde W(kr)$.  We carry out our analysis in
co-ordinate space where averaging kernels have exactly same volume but
somewhat different shapes. We do not expect this small variation in
the shapes to make a difference when the kernels have larger volumes.

One can then use the entropy deficit
$(H_{\theta\phi})_{max}-H_{\theta\phi}(r)$ of a distribution to
determine its linear bias on large scales where the density
fluctuations are smaller. On large scales, $P_{g}(k)=b^{2} P_{m}(k)$
and consequently the linear bias is given by,
\begin{eqnarray}
b=\sqrt{
  \frac{[(H_{\theta\phi})_{max}-H_{\theta\phi}(r)]_{g}}{[(H_{\theta\phi})_{max}-H_{\theta\phi}(r)]_{m}}}
&=&\sqrt{\frac{[a_{\theta\phi}(r)]_{g}}{[a_{\theta\phi}(r)]_{m}}}
\label{eq:biasval}
\end{eqnarray}
where the subscripts $g$ and $m$ corresponds to galaxy and mass
respectively.

The same argument also holds for the polar and azimuthal anisotropies
and one can use them independently to measure the linear bias for a
given galaxy distribution. We do not expect them to be different and
it would be interesting to measure and compare them. We analyze the
anisotropies in the galaxy distribution from the 2MRS following the
method outlined in this section and determine the linear bias
parameter.

It may be noted that one can use any spherical co-ordinates for this
analysis. In the present work, we use the galactic co-ordinates
$(l,b)$. Accordingly we replace $\theta$ and $\phi$ in the previous
definitions by $b$ and $l$ respectively.

\begin{figure*}
\resizebox{9cm}{!}{\rotatebox{-90}{\includegraphics{selfunc_mrs.ps}}}%
\resizebox{9cm}{!}{\rotatebox{-90}{\includegraphics{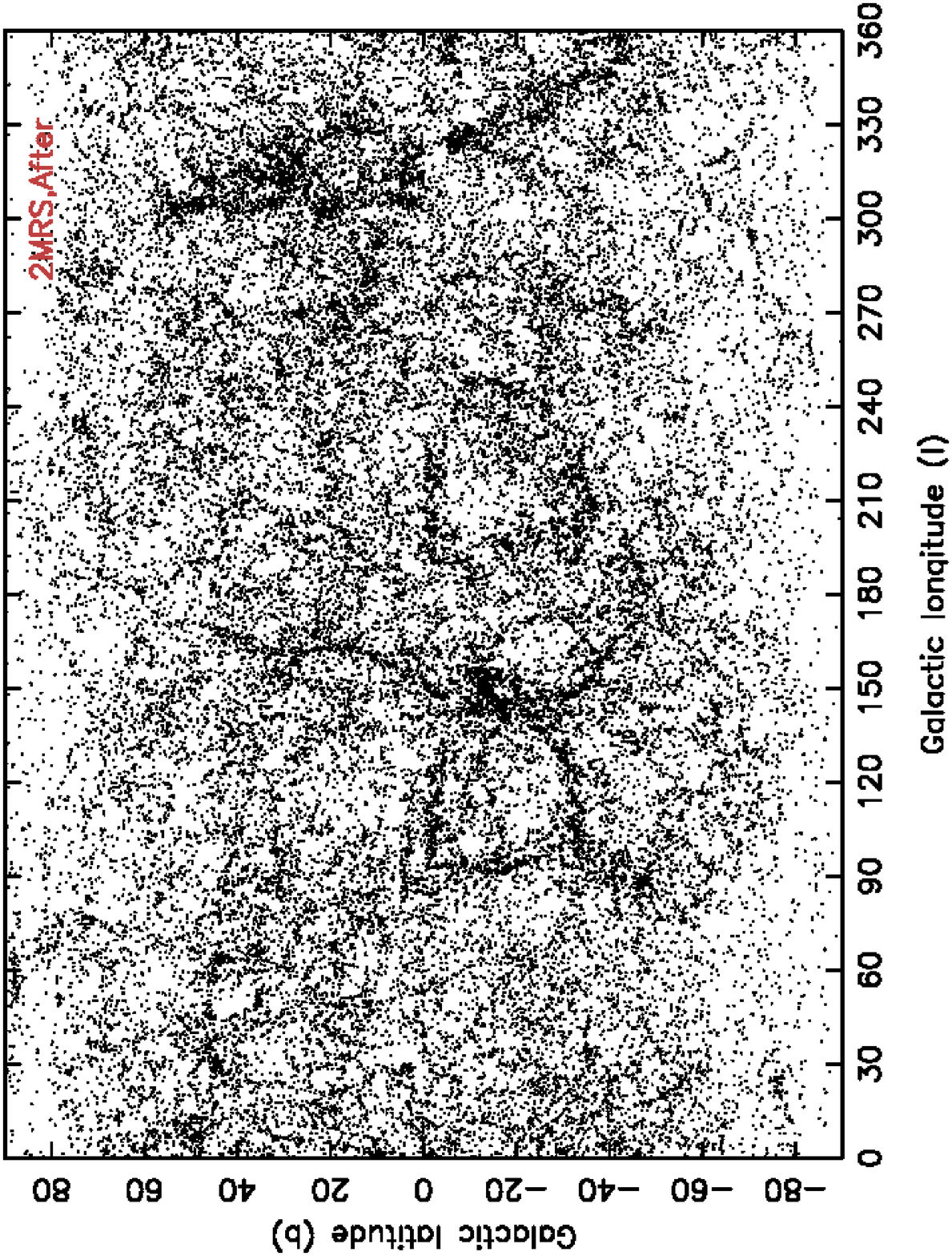}}}\\
\caption{ The left panel shows the redshift histogram in the 2MASS
  redshift survey (2MRS) along with the best fit (\autoref{eq:fit}) to
  it. The right panel shows the distribution of galactic coordinates
  of the 2MRS galaxies after the zone of avoidance is filled with
  cloned galaxies.}
  \label{fig:2mrs}
\end{figure*}

\begin{figure*}
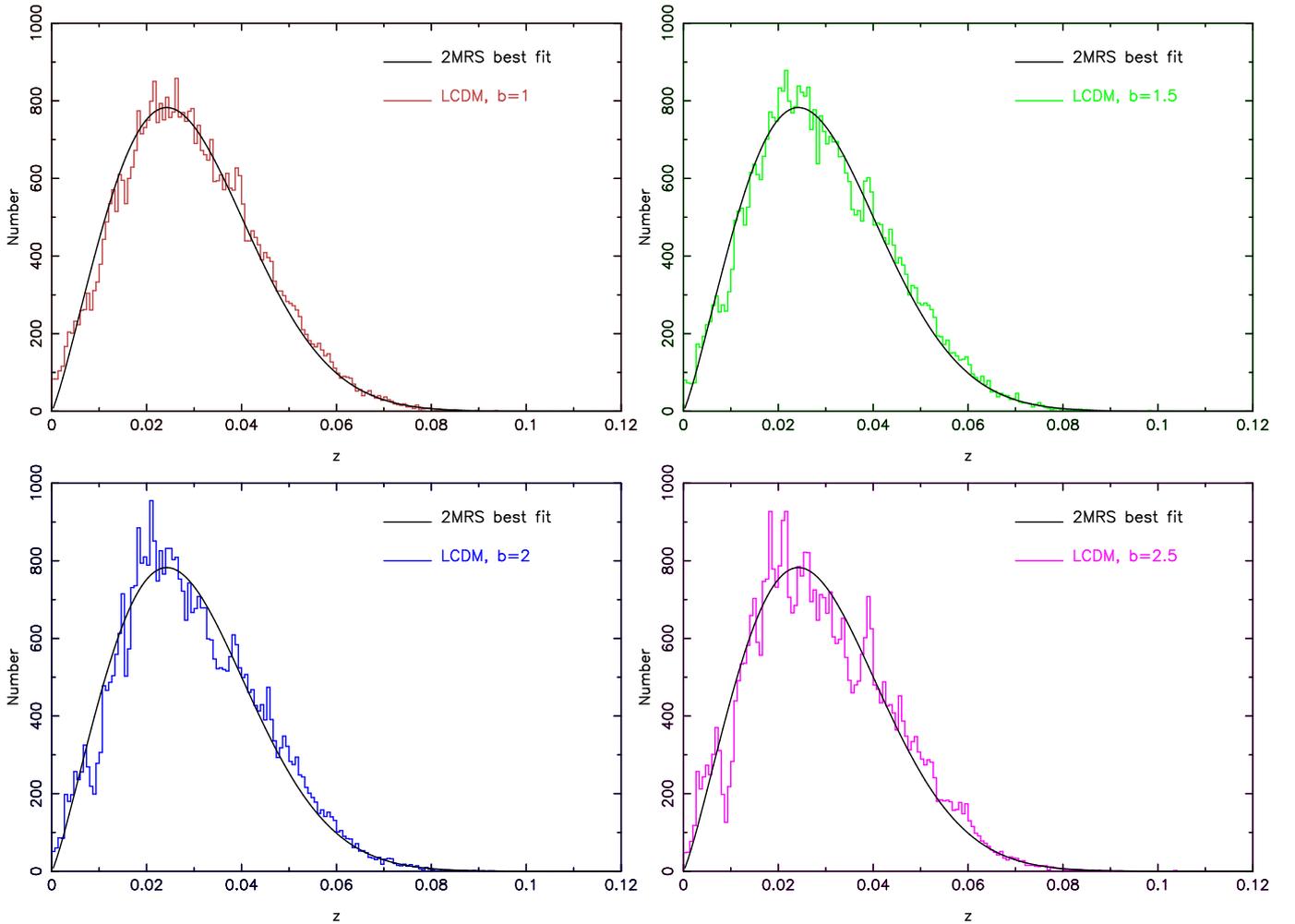

\resizebox{9cm}{!}{\rotatebox{-90}{\includegraphics{selfuncb1.ps}}}%
\resizebox{9cm}{!}{\rotatebox{-90}{\includegraphics{selfuncb1.5.ps}}}\\
\resizebox{9cm}{!}{\rotatebox{-90}{\includegraphics{selfuncb2.ps}}}%
\resizebox{9cm}{!}{\rotatebox{-90}{\includegraphics{selfuncb2.5.ps}}}\\
\caption{ Different panels of this plot show the redshift histograms
  for a simulated galaxy sample with different bias values along with
  the best fit to the 2MRS galaxy sample. The linear bias values of
  the respective samples are indicated in each panel. We use the best
  fit (\autoref{eq:fit}) to the 2MRS redshift distribution to simulate
  the mock galaxy catalogues for unbiased and biased distributions
  from N-body simulation of the $\Lambda$CDM model.}
  \label{fig:selfunc}
\end{figure*}

\begin{figure*}
\resizebox{9cm}{!}{\rotatebox{-90}{\includegraphics{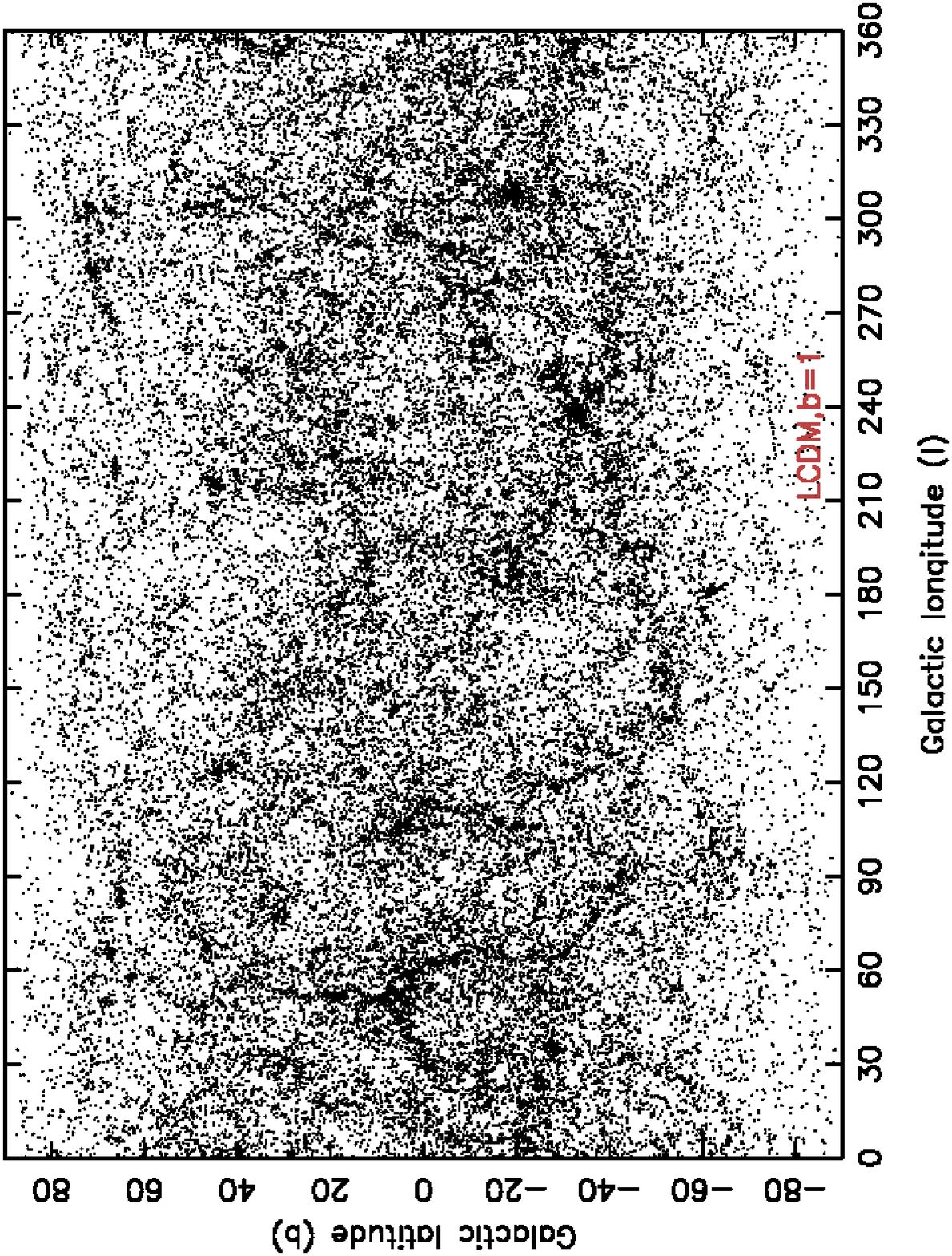}}}%
\resizebox{9cm}{!}{\rotatebox{-90}{\includegraphics{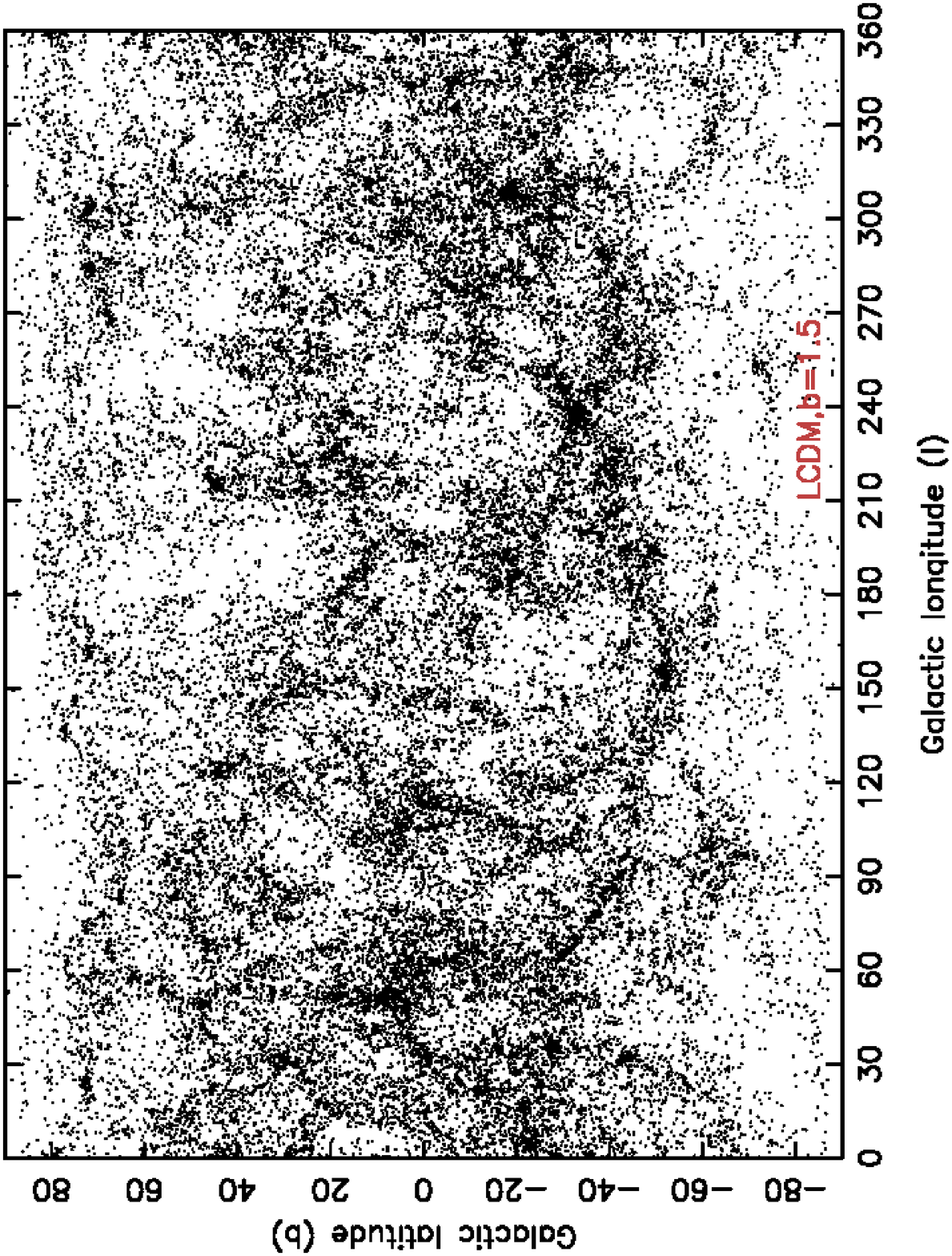}}}\\
\resizebox{9cm}{!}{\rotatebox{-90}{\includegraphics{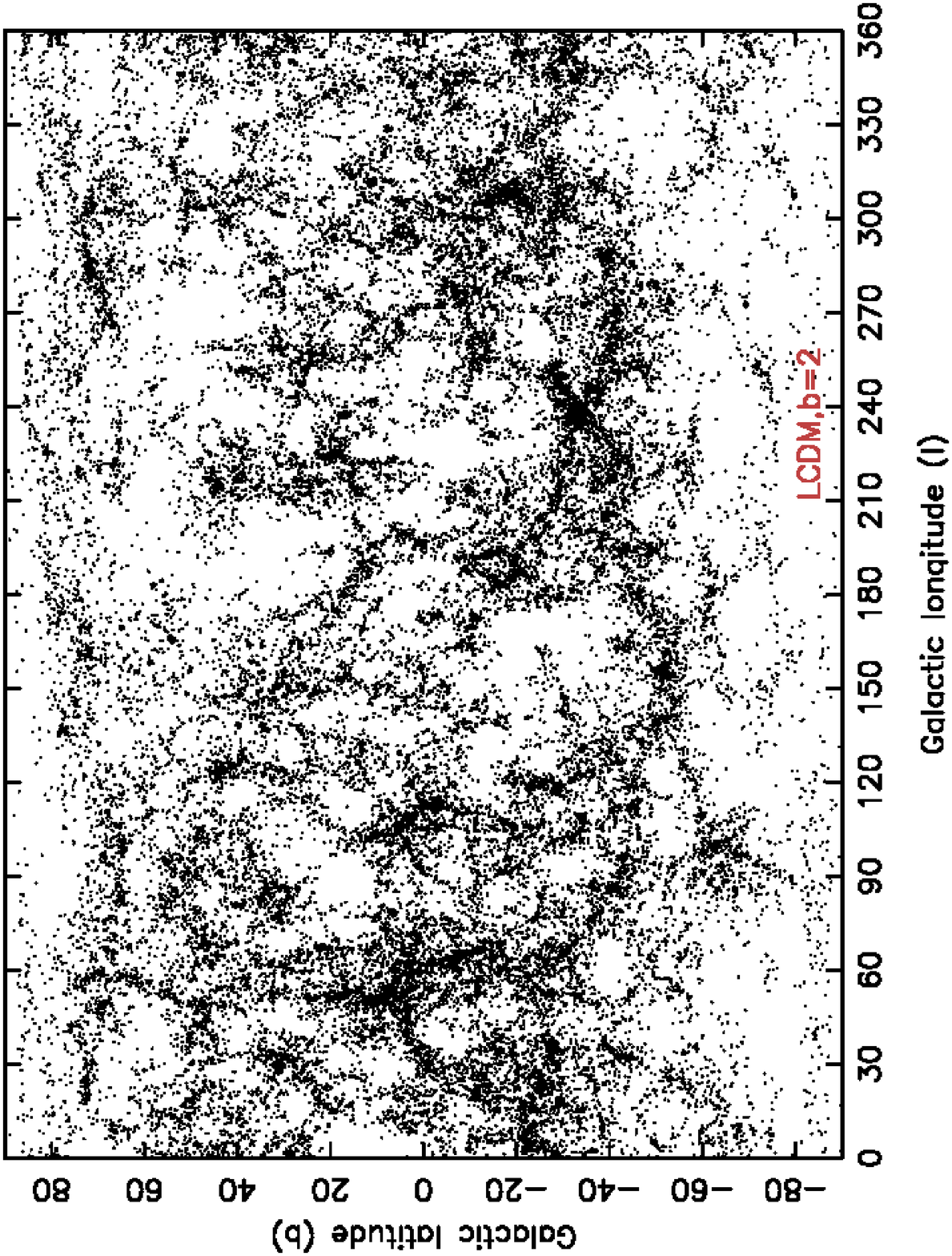}}}%
\resizebox{9cm}{!}{\rotatebox{-90}{\includegraphics{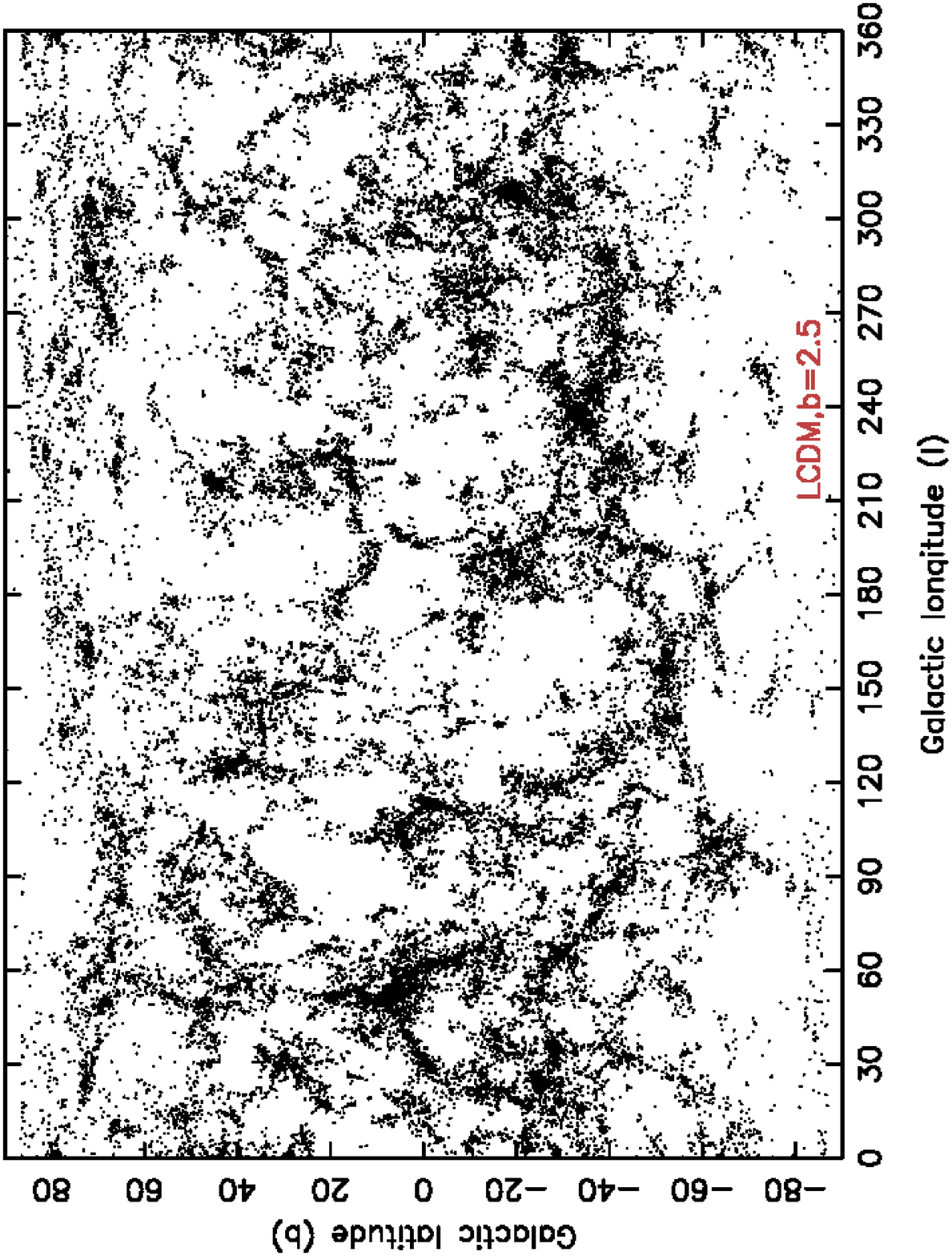}}}\\
\caption{ This shows the galactic coordinates in a mock 2MRS galaxy sample
  with different bias values as indicated in each panel.}
  \label{fig:dist}
\end{figure*}

\section{DATA}

\subsection{2MRS CATALOGUE}

The Two Micron All Sky Redshift Survey (2MRS) \citep{huchra} is an
all-sky redshift survey in the near infra-red wavelengths. The survey
is $97.6\%$ complete to a limiting magnitude of $K_{s}=11.75$ and
covers $91\%$ of the sky. It provides the spectroscopic redshifts of
$\sim 45,000$ galaxies in the nearby Universe. 2MRS selects the
galaxies with apparent infrared magnitude $K_{s} \leq 11.75$ and
colour excess $E(B-V) \leq 1$ in the region $|b| \geq 5^{\circ}$ for
$30^{\circ} \leq l \leq 330^{\circ}$ and $|b| \geq 8^{\circ}$
otherwise. \citet{huchra} rejected the sources which are of galactic
origin (multiple stars, planetary nebulae, HII regions) and discarded
the sources which are in regions of high stellar density and
absorption. The final 2MRS catalog by \citet{huchra} contain $43,533$
galaxies. We restrict our sample to $z\leq 0.12$ beyond which there
are a very few galaxies. This redshift limit is used to simulate the
mock catalogues for the 2MRS. We use this 2MRS flux limited sample
which contains $43,305$ galaxies.

To construct mock catalogues for the 2MRS flux limited sample we
first model the redshift distribution using a parametrized fit 
\citep{erdogdu2, erdogdu1} given by,
\begin{equation}
\frac{dN(z)}{dz}= A \,z^{\gamma} \, \exp[-{\big(\frac{z}{z_{c}}\big)}^\alpha]
\label{eq:fit}
\end{equation} 
We calculate the redshift distribution in the 2MRS using uniform bin
size of $200 \, km/s$ and then fit it with \autoref{eq:fit} using the
nonlinear least-squares method (Marquardt-Levenberg algorithm). Each
point in the data are assigned equal weights. We find the values of
the best fit parameters to be $A = 116000 \pm 5100$, $\gamma = 1.188
\pm 0.093$, $z_{c} = 0.031 \pm 0.002$ and $\alpha = 2.059 \pm
0.149$. The redshift histogram in the 2MRS along with the best fit
(\autoref{eq:fit}) curve is shown in the left panel of
\autoref{fig:2mrs}. It may be noted that dividing \autoref{eq:fit} by
the total number of galaxies gives the probability of detecting a
galaxy at redshift $z$.

We would like to have a galaxy distribution over full-sky for our
analysis. This requires us to artificially fill the Zone of Avoidance
(ZOA), the region near the Galactic plane which is obscured due to the
extinction by Galactic dust and stellar confusion. We randomly select
galaxies from the unmasked region and then place them at random
locations in the masked area so as to have the same average density in
the masked and unmasked region \citep{lyndenbell}. We clone $4,375$
galaxies to fill the ZOA and after carrying out the cloning procedure,
finally we have $47,680$ galaxies in our 2MRS sample. The distribution
of the galactic coordinates of galaxies in the 2MRS after filling the
ZOA is shown in the right panel of \autoref{fig:2mrs}. We construct $30$
jackknife samples from the 2MRS data each containing $35,000$ galaxies.

\begin{figure*}
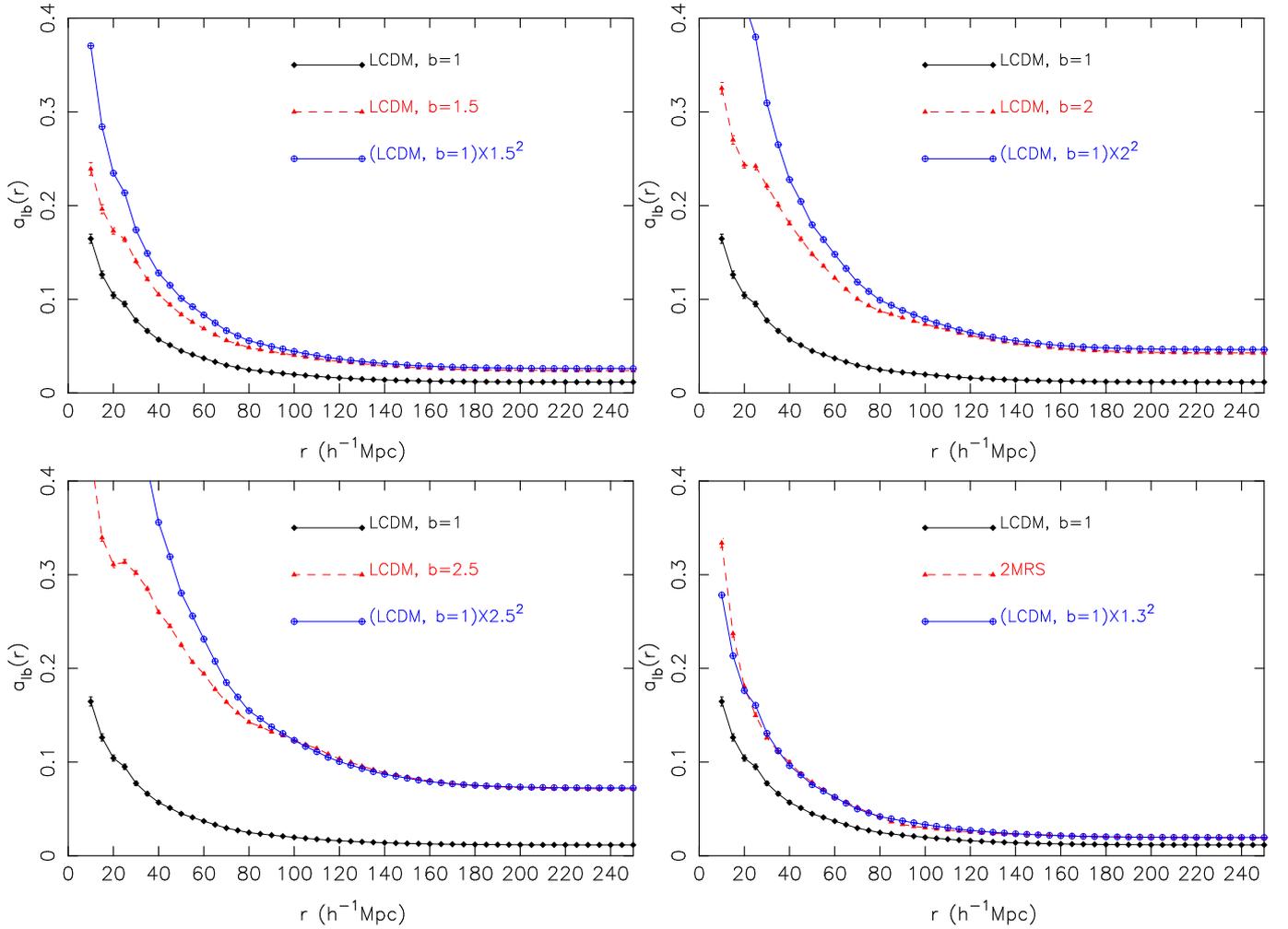

\resizebox{9cm}{!}{\rotatebox{-90}{\includegraphics{plot1_rr.ps}}}%
\resizebox{9cm}{!}{\rotatebox{-90}{\includegraphics{plot2_rr.ps}}}\\
\resizebox{9cm}{!}{\rotatebox{-90}{\includegraphics{plot3_rr.ps}}}%
\resizebox{9cm}{!}{\rotatebox{-90}{\includegraphics{plot4_rr.ps}}}\\
\caption{The top left, top right and bottom left panel show that the
  radial anisotropies in the simulated galaxy samples can be obtained
  by scaling the radial anisotropies in the dark matter distribution
  by $b^{2}$ where $b$ is the linear bias parameter of the simulated
  galaxy sample. The bottom right panel shows that the radial
  anisotropies in the 2MRS galaxy samples is well represented by the
  radial anisotropies expected in a galaxy distribution in
  $\Lambda$CDM model with linear bias $b=1.3$. The radial anisotropies
  for the simulated samples shown in each panel are the mean
  anisotropies obtained from $30$ mocks in each case. The $1\sigma$
  errorbars shown in each panel are obtained from $30$ mock catalogues
  for the N-body simulations and $30$ jackknife samples for the 2MRS
  data.}
  \label{fig:rr}
\end{figure*}

\begin{figure*}
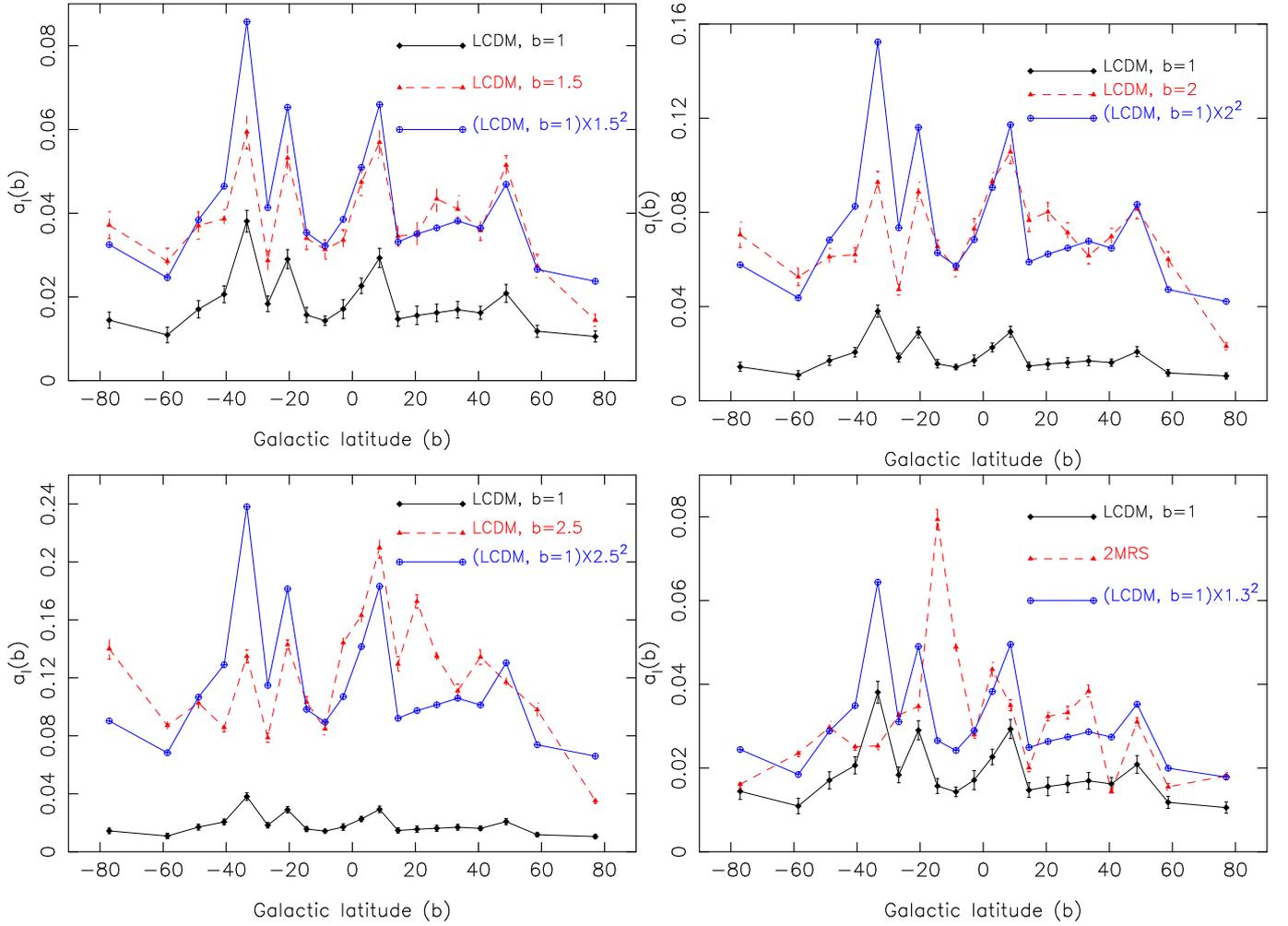

\resizebox{9cm}{!}{\rotatebox{-90}{\includegraphics{plot1_theta.ps}}}%
\resizebox{9cm}{!}{\rotatebox{-90}{\includegraphics{plot2_theta.ps}}}\\
\resizebox{9cm}{!}{\rotatebox{-90}{\includegraphics{plot3_theta.ps}}}%
\resizebox{9cm}{!}{\rotatebox{-90}{\includegraphics{plot4_theta.ps}}}\\
\caption{Same as \autoref{fig:rr} but for polar anisotropies.}
  \label{fig:bb}
\end{figure*}

\begin{figure*}
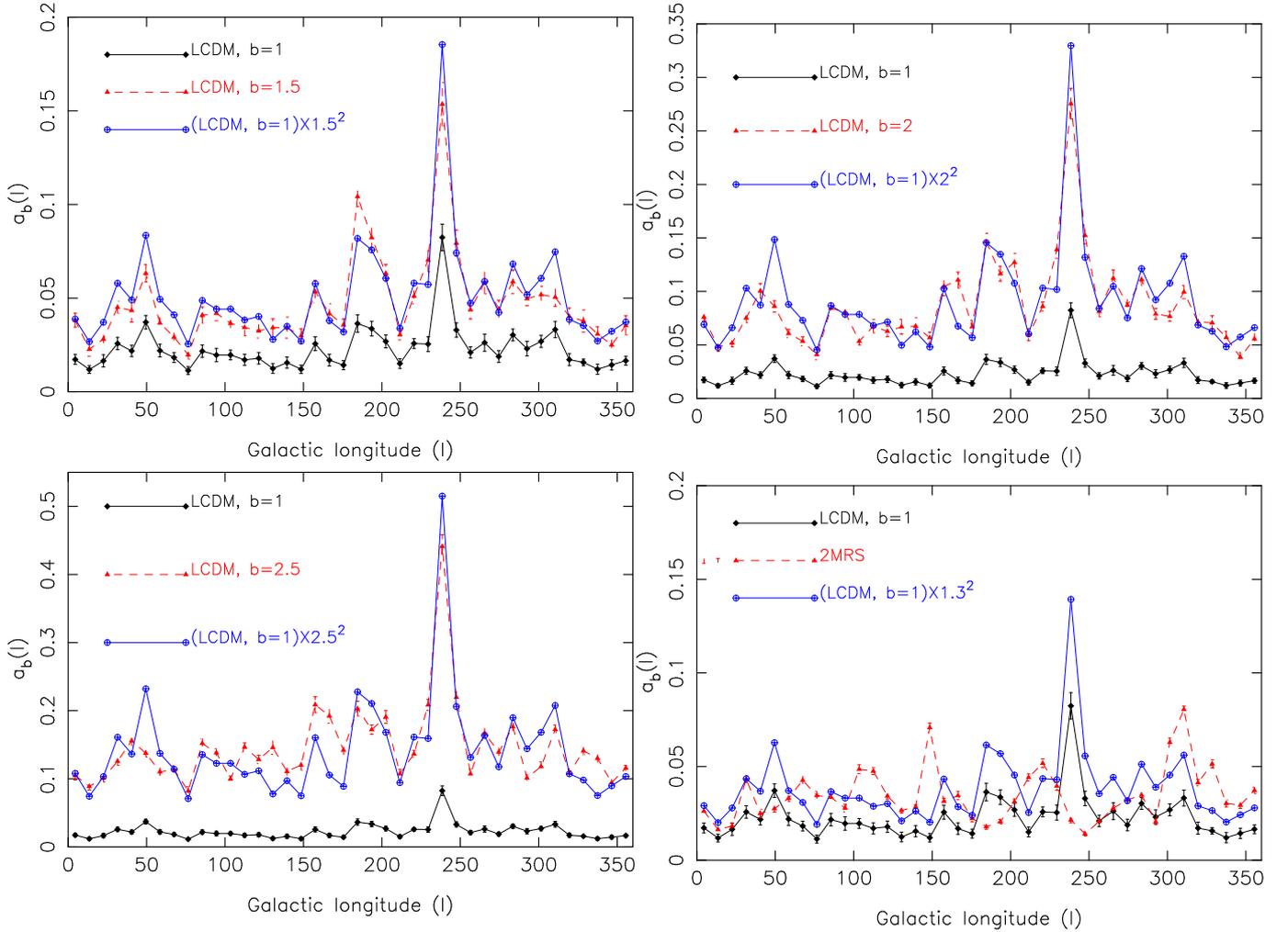

\resizebox{9cm}{!}{\rotatebox{-90}{\includegraphics{plot1_phi.ps}}}%
\resizebox{9cm}{!}{\rotatebox{-90}{\includegraphics{plot2_phi.ps}}}\\
\resizebox{9cm}{!}{\rotatebox{-90}{\includegraphics{plot3_phi.ps}}}%
\resizebox{9cm}{!}{\rotatebox{-90}{\includegraphics{plot4_phi.ps}}}\\
\caption{Same as \autoref{fig:rr} but for azimuthal anisotropies.}
  \label{fig:ll}
\end{figure*}

\subsection{MOCK CATALOGUES FROM N-BODY SIMULATION}
We use a Particle-Mesh (PM) N-body code to simulate the present day
distributions of dark matter in the $\Lambda$CDM model in a comoving
volume of $[921.6 h^{-1} {\rm Mpc}]^3$. We use $256^3$ particles on a
$512^3$ mesh and the following cosmological parameters:
$\Omega_{m0}=0.31$, $\Omega_{\Lambda0}=0.69$, $h=0.68$,
$\sigma_{8}=0.81$ and $n_{s}=0.96$ \citep{adeplanck3} are used in the
simulation. In the current paradigm, the galaxies are believed to form
at the peaks of the density field. We implement a simple biasing
scheme \citep{cole} where the galaxies are allowed to form only in
those peaks where the overdensity exceeds a certain density
threshold. One can vary the threshold in this sharp cut-off biasing
scheme to generate galaxy distributions with different bias values. We
determine the linear bias parameter $b$ for these samples as,
\begin{equation}
b=\sqrt{ \frac {\xi_g(r)}{\xi_{m}(r)}}
\end{equation}
where $\xi_g(r)$ and $\xi_{m}(r)$ are the two-point correlation
functions for the galaxy and dark matter distribution respectively.
We generate the distributions for three different bias values $b=1.5$,
$b=2$ and $b=2.5$.

We construct a set of mock 2MRS catalogues from the unbiased and
biased distributions. The function given in \autoref{eq:fit} has the
maxima at $z=z_{c}
(\frac{\gamma}{\alpha})^\frac{1}{\alpha}$. Substituting the best fit
values of the parameters $A$, $\gamma$, $z_{c}$ and $\alpha$, we find
that the maximum probability of finding a galaxy in the 2MRS sample is
at $z_{max}=0.0242$. One can then calculate the maximum probability
$P_{max}$ from \autoref{eq:fit} using the values of $z_{max}$ and the
best fit parameters. To simulate the 2MRS mock catalogues from the
N-body simulation and the biased distributions, we treat the particles
as galaxies and place an observer at the center of the box. We map the
galaxies to redshift space using their peculiar velocities. We
randomly choose a galaxy within the redshift range $0 \leq z \leq
0.12$ and calculate the probability of detecting this galaxy using
\autoref{eq:fit}. We also randomly choose a probability value in the
range $0 \leq P(z) \leq P_{max}$. If the calculated probability is
larger than the randomly selected probability then we retain the
randomly selected galaxy in our sample. This process is repeated until
we have $47,680$ galaxies in the mock sample. We extract $30$ mock
samples for each bias values. We show the all sky distribution of the
galactic coordinates for a mock 2MRS sample with different bias values
in \autoref{fig:dist}.

\section{Results and Conclusions}
In \autoref{fig:rr} we compare the radial anisotropies $a_{lb}(r)$ in
the simulated mock biased galaxy samples with that from the mock
samples from the dark matter distribution in the $\Lambda$CDM
model. The top left panel, top right panel and the bottom left panel
of \autoref{fig:rr} show the comparisons for linear bias values
$b=1.5$, $b=2$ and $b=2.5$ respectively. We see in each of these
panels that scaling the radial anisotropy with $b^{2}$ where $b$ is
the linear bias parameter of the simulated galaxy sample, reproduces
the actual radial anisotropies observed in the respective galaxy
samples on large scales. However this scaling shows a large deviation
on small scales which gradually decreases and finally merges with the
observed radial anisotropies in the biased samples beyond a length
scale of $90 \hmpc$. We do not expect the linear biasing to hold on
small scales. On small scales, the differences result from the
non-linearities present on those scales due to the gravitational
clustering. The contributions from the higher order moments of the
probability distribution in \autoref{eq:shannon3} are not negligible
on smaller scales and the bias values obtained by using
\autoref{eq:biasval} are expected to deviate from its actual value.
Eventually the assumption of linear bias may prevail on some larger
scale and the \autoref{eq:biasval} can faithfully recover the linear
bias values only on a scale where the non-linearity becomes
negligible. In the bottom right panel of \autoref{fig:rr} we compare
the radial anisotropies in the 2MRS galaxy sample with that expected
from the unbiased $\Lambda$CDM model. Interestingly, when we scale the
radial anisotropies in the unbiased $\Lambda$CDM model by $1.3^{2}$ we
find that it nicely represents the radial anisotropies observed in the
2MRS galaxy sample for nearly the entire length scales beyond $20
\hmpc$. This indicates that the non-linearity becomes less important
in the 2MRS galaxy sample beyond a length scale of $20 \hmpc$. It is
also interesting to note that though the radial anisotropy in all the
biased galaxy distributions decreases with increasing length scales,
they reach a plateau at different length scales. We note that for
$b=1$, $b=1.5$, $b=2$ and $b=2.5$ the plateaus are reached at $90
\hmpc$, $130 \hmpc$, $150 \hmpc$ and $170 \hmpc$ respectively. This
indicates that the signatures of anisotropy may persist up to different
length scales depending on the bias of the galaxy distribution.

\begin{table*}{}
\caption{This shows the linear bias values estimated from the polar
  and azimuthal anisotropies for the simulated samples and the 2MRS
  sample. We calculate the linear bias values using
  \autoref{eq:biasval} but with the average polar and azimuthal
  anisotropies measured from the $30$ samples in each case. We average
  the bias measurements from polar and azimuthal anisotropies over
  different latitudes and longitudes respectively. The errors quoted
  with the bias values in the table are the standard errors.}
\label{tab:biaslb}
\begin{tabular}{|c|c|c|c|}
\hline
Sample & Bias from $a_{l}(b)$ & Bias from $a_{b}(l)$\\
\hline
$\Lambda$CDM,$b=1$&$1$&$1$\\
$\Lambda$CDM,$b=1.5$&$1.45\pm0.026$&$1.44\pm0.017$\\
$\Lambda$CDM,$b=2$&$1.96\pm0.051$&$1.96\pm0.033$ \\
$\Lambda$CDM,$b=2.5$&$2.57\pm0.089$&$2.6\pm0.053$ \\
2MRS&$1.31\pm0.067$&$1.29\pm0.055$ \\
\hline
\end{tabular}
\end{table*}

Our scheme maintains equal area for all the pixels by uniformly
binning $\cos\theta$ and $\phi$. This causes the shapes of the pixels
to vary across different parts of the sky. These variations may
contribute to the anisotropies measured in our scheme. To assess this
we carry out some tests with HEALPix \citep{gorski1,gorski2} which
uses equal area and nearly same shape for all the pixels. We calculate
the radial anisotropy in the same datasets using NSide$=8$ in HEALPix
which provides a total $768$ pixels on the sky. It may be noted that
we use $m_{b}=20$ and $m_{l}=40$ in our scheme which results into a
total $800$ pixels. We find that HEALPix pixelization gives exactly
the same radial anisotropy as measured in our scheme \citep{pandey17}.

We compare the polar and azimuthal anisotropies in the biased and
unbiased samples in the top left, top right and bottom left panels of
\autoref{fig:bb} and \autoref{fig:ll} respectively. We notice that a
scaling similar to \autoref{fig:rr} also applies here despite the fact
that a smaller number galaxies are used to compute the anisotropies at
each $b$ and $l$. It may be noted that the peaks and troughs in the
polar and azimuthal anisotropy curves for the simulated samples appear
nearly at the same $l$ and $b$ values as the biased distributions are
produced from the same unbiased distribution. But if we compare these
results with that from a galaxy distribution, we do not expect this to
happen as they represent two different statistical realizations of the
density field. We find that the \autoref{eq:biasval} can be also used
effectively with polar and azimuthal anisotropies to recover the
linear bias parameter of the biased galaxy samples
(\autoref{tab:biaslb}). We consider the polar and azimuthal
anisotropies estimated from $30$ samples in each case to measure the
linear bias values using \autoref{eq:biasval} respectively at each
latitude ($b$) and longitude ($l$). We estimate the average linear
bias values and their standard errors by combining the bias
measurements over different latitudes and longitudes and list them in
\autoref{tab:biaslb}.

In \autoref{tab:biaslb} we see that the linear bias values recovered for
the simulated galaxy samples are quite close to their actual bias
values. When we apply the same method to estimate the linear bias of
the 2MRS galaxy sample, we get $b=1.31$ from polar anisotropy and
$b=1.29$ from azimuthal anisotropy. It is interesting to note that we
get nearly the same bias value $b\sim1.3$ from radial, polar and
azimuthal anisotropies. One can also determine the relative bias
parameter between any two galaxy distributions using the same method.

We also test the applicability of this method to mock samples where
the radial selection function is uniform. We randomly extract $10^5$
particles from $10$ spherical regions of radius $200 \hmpc$ from each
of the biased and unbiased distributions and repeated the analysis. We
find that one can recover the linear bias of the simulated galaxy
samples following the same method presented in this work. This
suggests that the same method can be applied to determine the linear
bias parameter of the volume limited sample from different galaxy
surveys.

 It may be noted that the computation of the two-point correlation
 function and power spectrum scales as $O(N^{2})$ where $N$ is the
 number of galaxies in the sample. So the computational requirements
 scales very fast with the size of the sample. Use of tree algorithms
 or FFT can reduce this scaling to $O(N \log N)$
 \citep{szapudi1,pen,szapudi2}. Interestingly the method presented in
 this work requires a scaling of only $O(N)$ and hence it is
 computationally least expensive among all the other existing methods
 for the determination of linear bias.

We finally note that a combined study of the radial, polar and
azimuthal anisotropies in the galaxy distribution provides a powerful
new alternative to measure the linear bias parameter from galaxy
distributions.

\section{Acknowledgement}
The author thanks an anonymous reviewer for the valuable comments and
suggestions. The author would like to thank the 2MRS team for making
the data public. The author acknowledges financial support from the
SERB, DST, Government of India through the project EMR/2015/001037. I
would also like to acknowledge IUCAA, Pune and CTS, IIT, Kharagpur for
providing support through associateship and visitors programme
respectively.

\bsp	
\label{lastpage}
\end{document}